\documentclass[twocolumn,epjc3]{svjour3}
\smartqed  
\RequirePackage{graphicx}
\usepackage{color}

\begin{document}

\title{Energetics and optical properties of $6$-dimensional rotating black hole in pure Gauss-Bonnet gravity}

\author{
Ahmadjon Abdujabbarov\thanksref{1,a,b} \and
Farruh Atamurotov\thanksref{2,a,c} \and
Naresh~Dadhich\thanksref{3,d,e} \and
Bobomurat Ahmedov\thanksref{4,a,b} \and
Zden\v{e}k Stuchl\'{i}k\thanksref{5,f} \and}

\institute{
Institute of Nuclear Physics,~Ulughbek, Tashkent 100214,
Uzbekistan\label{a}\and Ulugh Beg Astronomical Institute, Astronomicheskaya 33,
Tashkent  100052, Uzbekistan\label{b}\and
Inha University in Tashkent, Tashkent 100170, Uzbekistan\label{c}
\and
Inter-University Centre for Astronomy and Astrophysics, Post Bag 4, Ganeshkhind,
Pune 411 007, India\label{d}\and Centre for Theoretical Physics,
Jamia Millia Islamia, New Delhi 110025, India\label{e}
\and
Institute of Physics, Faculty of Philosophy and Science,
Silesian University in Opava, Bezrucovo nam. 13, Opava, Czech Republic\label{f} }

\thankstext{1}{\emph{e-mail:} ahmadjon@astrin.uz}
\thankstext{2}{\emph{e-mail:} farruh@astrin.uz}
\thankstext{3}{\emph{e-mail:} nkd@iucaa.ernet.in}
\thankstext{4}{\emph{e-mail:} ahmedov@astrin.uz}
\thankstext{5}{\emph{e-mail:} zdenek.stuchlik@fpf.slu.cz}

\maketitle

\begin{abstract}

We study physical processes around a rotating black hole in pure Gauss-Bonnet (GB) gravity.
In pure GB gravity,  gravitational potential has slower fall off  as compared to
the corresponding Einstein potential in the same dimension. It is therefore expected that the energetics of
pure GB black hole would be weaker, and our analysis bears out that the efficiency of  energy extraction by
Penrose process  is increased to   $25.8\%$  and  particle acceleration  is increased to   $55.28\%$,  and optical
shadow of the black hole is decreased. These are the distinguishing in principle observable features of pure GB black hole.

\end{abstract}

\section{Introduction}

From all the generalizations of Einstein
gravity what distinguishes the Lovelock gravity is its remarkable
property that the equation of motion always remains second order.
This happens despite the Lagrangian being polynomial
in Riemann curvature. It is the underlying differential geometric
structure that is responsible for this unique remarkable property,
discovered by Lovelock~\cite{lovelock}. Lovelock action
$S$ which is a homogeneous polynomial of degree $N$ in Riemann
curvature reads as %
\begin{equation}
\label{action} S=\int \sqrt{-g}\left(\sum_{N}
\lambda_{{N}} {\cal L}_{{N}} \right)d^D x \ ,
\end{equation}
where the Lagrangian is given by
\begin{eqnarray}
{\cal L}_N=\frac{1}{2^N}\delta^{\alpha_1...\alpha_{_N}\beta_1...\beta_{_N}}_{\gamma_1..\gamma_{_N}\mu_1..\mu_{_N}}R^{\gamma_1\mu_1}_{\ \ \ \ \ \alpha_1\beta_1}...R^{\gamma_{_N}\mu_{_N}}_{\ \ \ \ \ \alpha_{_N}\beta_{_N}}\ ,
\end{eqnarray}
and  $\lambda_{N}$ are coupling constants.
In this
action $N = 0, 1, 2, ...$ respectively correspond to $\Lambda$,
the usual Einstein-Hilbert, Gauss-Bonnet and so on. However higher
order Lovelock terms in action make non-zero contribution only in
dimensions $>4$. That is Lovelock is the natural higher dimensional
generalization of Einstein gravity. \\

It is well-known that particle physics theories, in particular string
theory, call for higher dimensions as demanded by physical symmetries.
Besides one of us had also articulated some
purely classical considerations for higher dimensions
\cite{dadhich04,dadhich05,dadhich11}. One of the most convincing
arguments goes as follows \cite{dadhich11}. It is customary to consider high energy
effects of a theory by taking higher power of the basic variable.
In the case of gravity, the basic physical entity is Riemann
curvature and hence to take into account high energy effects, we
should include higher powers of it in the action. However if we demand that the
equation of motion should not change its second order character,
then we are uniquely led to Lovelock action which is pertinent only in
higher dimensions. That is high energy effects of gravity are accessible
only in higher dimensions. Thus it makes a good case for studying gravity in higher
dimensions. \\

Note that in the above action, there is sum over $N$ and each term has a dimensionful
coupling constant which cannot be determined because experiment can determine only one
constant. To make problem tractable, it was assumed, for obtaining dimensionally continued black holes \cite{banados94}, that all the couplings were given
in terms of the unique ground state $\Lambda$. On the other hand, there is a strong
case made out for pure Lovelock gravity \cite{dadhich12a,Dadhich12c} where there is only one $N$th order term in the action with $\Lambda$. That is,  it does not
include even the usual Einstein-Hilbert term. This has remarkable
unique property that gravitational dynamics is distinguished in all
odd and even dimensions \cite{Dadhich12c}. In all odd $D=2N+1$ dimensions, gravity is kinematic meaning $N$th order Lovelock analogue Riemann tensor
vanishes whenever the corresponding Ricci tensor vanishes \cite{Camanho15}. This of course includes the usual $3$ dimension for which Riemann tensor is entirely given in terms of
Ricci tensor showing kinematic property of Einstein gravity for $N=1$. It is thus a universal feature of pure Lovelock gravity. We will therefore adhere to pure
Lovelock generalization for exploring gravity in higher dimensions. \\

Rotating black holes are by far the most exciting objects as they
offer a rich arena of interesting physical processes of
astrophysical importance in the usual four spacetime  dimensions.
These include high energy exotic objects like quasars and active
galactic nuclei (AGNs) with their energetic jets,  energy
extraction processes by Penrose process \cite{Penrose69} and its
magnetic version
\cite{wagh85,wagh86erratum,wagh85a,Part,dadhich12b,Koide14}  and
Blandford-Znajek mechanism \cite{Blandford1977} and  particle
acceleration \cite{Banados09} as well as optical shadow of black hole, see,
e.g.,
\cite{Virbhadra1,Virbhadra2,Bozza2005,Bambi10,Grenzebach2014,Falcke2013b,Abdujabbarov15}.
The possibility to obtain ultra-high energy particles and the appearance of Keplerian discs orbiting Kerr superspinars have been studied in~\cite{Stuchlik13,Stuchlik10}.
Keeping in view physical
richness and astrophysical  significance
of rotating black holes, it would be pertinent to probe these
interesting properties for higher dimensional rotating black holes.
As argued above, in higher dimensions,  pure Lovelock gravity
enjoys an unique special position in view of its universal
characteristics for all odd and even dimensions. It would
therefore be pertinent to study interesting physical and
astrophysical phenomena for a pure GB rotating black hole in six
dimension. Apart from astrophysical motivation for this study,
there is a fundamental question of what shape  gravitational
dynamics takes in higher dimensions? For instance in Einstein
gravity,  gravitational field becomes stronger as dimension
increases which implies that there can exist no bound orbits in
dimension greater than four while for pure Lovelock gravity, it
becomes weaker with dimension and hence bound orbits continue to
exist in all even dimensions \cite{dadhich13}. \\

 Very recently, a rotating black hole metric has been
obtained ~\cite{dadhich2013} to describe an analogue of Kerr black
hole in  pure GB gravity in six dimensions. Though it is not an
exact solution of the pure Lovelock vacuum equation, it has all the
desirable and expected features and it
satisfies the equation in the leading order. In this paper, we
wish to study energetics of pure GB rotating black hole by
employing this metric. Our study yields results which are in tune
with the general properties of a rotating black hole and hence it
further lends support to the metric for its viability. \\

Note that Einstein gravity is vacuous in
two, kinematic in three and becomes dynamic in four dimensions. and it is
pure Lovelock of order $N=1$. The universalization of this general gravitational feature means
gravity should respectively be vacuous, kinematic and dynamic for
$D = 2N, 2N+1, 2N+2$. This uiniquely picks out pure Lovelock gravity; i.e. pure Lovelock should be the
gravitational equation in higher dimensions. That is in all odd and even $D=2N+1, 2N+2$ dimenisons gravitational
dynamics should be similar~\cite{dadhich15}. This is what has been verified in various situations; for instance black hole
entropy is always proportional to square of horizon radius \cite{dadhich12a,Dadhich12c} and bound orbits around static black hole exist
only for the pure Lovelock gravity in all even $D=2N+2$ dimensions~\cite{dadhich13}. This motivates us to examine this general feature
in all possible situations and that is what we wish to do it in this paper. We shall therefore study all the usual physical
processes like energy extraction, Hawking radiation, optical shadow and particle acceleration for a pure GB rotating black hole
\cite{dadhich2013} and compare them to that of a rotating black hole in the usual four dimensional spacetime.
This is the primary aim of the paper. \\

The paper is arranged as follows: the rotating black hole metric
is discussed in Sec.~\ref{metric1} while in Sec. \ref{second}, we
analyze the geodesic equations for circular orbits which is followed by
discussion of optical shadow of black hole in Sec. \ref{third}. Next we study
energy extraction processes through BSW effect and Penrose process. We conclude with a
discussion. Throughout the paper, we use a system of units in which the GB coupling constant and
velocity of light are set to unity.

\section{Pure GB rotating black hole metric}\label{metric1}

We wish to consider a pure Gauss-Bonnet rotating black hole in the
critical $6$ dimension (in odd $d=5$ GB gravity is kinematic; i.e.
GB flat and it becomes dynamic in even $d=6$ \cite{Dadhich12c}).
There does not exist an exact solution of
pure GB vacuum equation for an axially symmetric spacetime
representing a rotating black hole. This is simply because the
equations are very formidable to handle. However for Einstein
gravity in $4$ dimension there is the well known Newman-Janis
algorithm for converting a static black hole into a rotating one
without solving the equation. This may however not be applicable
for GB gravity and in higher dimension \cite{Hansen13}. Secondly one of us
\cite{dadhich13b} has recently obtained the Kerr metric by appealing to
the two simple physical considerations. One, it should incorporate
Newton's law in the first approximation, and second, since one is free to choose affine parameter for a null curve and hence the radial coordinate is chosen as
affine parameter for radially falling photon. For this, one begins with
an appropriate spatial geometry which
has ellipsoidal symmetry and then implement these two common sense inspired
physical considerations, and what results is the metric \cite{dadhich2013} considered here. This
can describe a rotating black hole, though it is not an exact solution of
pure GB vacuum equation, yet it has all the features of the usual Kerr metric. It
however does satisfy the equation in the leading order.
It has all the characteristics of a rotating black hole in existence of ergosphere, and the right limits; for $a=0$ it reduces to pure GB static black hole while $M=0$ leads to flat space. It is therefore perfectly appropriate
metric for studying a rotating black hole in pure GB gravity. We shall thus employ this rotating black hole metric for studying its various physical properties.

The stationary axisymmetric metric for pure GB rotating black hole in the standard Boyer-Lindquist
coordinates  reads as \cite{dadhich2013}

\begin{eqnarray}
ds^2 &=& -\frac{\Delta}{\Sigma}\bigg[dt-a^2 \sin^2\theta d\phi \bigg]^{2}+\frac{\Sigma}{\Delta}dr^2 \nonumber\\
&&  + \Sigma d\theta^{2}+\frac{\sin^2\theta}{\Sigma}
\bigg[(r^2+a^2)d\phi-a dt \bigg]^2   \nonumber\\
&& + r^2 \cos^2\theta \big[d\psi^2 + \sin^2\psi d\chi^2 \big] \  ,
\label{metric}
\end{eqnarray}
with
\begin{eqnarray}\label{delta}
\Delta&=& r^2+a^2-M^{1/2}r^{3/2}\ ,\nonumber\\
\Sigma&=& r^2+a^2\cos^2\theta \ ,
\end{eqnarray}
where $M$ and $a$ have usual meaning of the total mass and specific angular momentum.

The spacetime (\ref{metric}) has  horizon when $t= const $
becomes null; i.e. $\Delta=0$ which has the following four roots
\begin{eqnarray}
r_{+,-}&=&\frac{M}{4}+\frac{C}{4\sqrt{3}} \nonumber \\ &&
\pm   \sqrt{\frac{3 M^2}{16}-a^2-\frac{C^2}{192} +\sqrt{3}\frac{
M^3-8a^2M}{8C}}\ , \label{r12} \\
r_{3,4}&=&\frac{M}{4}-\frac{C}{4\sqrt{3}} \nonumber \\ &&
\pm \sqrt{\frac{3 M^2}{16}-a^2-\frac{C^2}{192} -\sqrt{3}\frac{M^3-8a^2M}{8C}}\label{r34}\ ,
\nonumber\\
\end{eqnarray}
with
\begin{eqnarray}
C^2&=&3 M^2-16 a^2+\frac{64a^4}{A}+4A\ , \nonumber\\
A^3&=&\frac{27}{2}a^4M^2-64a^6+3\sqrt{3}a^4M D \ , \nonumber\\
D&=& \sqrt{\frac{27}{4}M^2-64 a^2} \ . \nonumber
\end{eqnarray}

The function under the square root in expression (\ref{r34})
$$
F_1(a,M)={\frac{3 M^2}{16}-a^2-\frac{C^2}{192} -\sqrt{3}\frac{M^3-8a^2M}{8C}}
$$
is always negative and consequently $r_{3,4}$ is not real, while
the function under the square root in expression (\ref{r12})
$$
F_2(a,M)={\frac{3 M^2}{16}-a^2-\frac{C^2}{192} +\sqrt{3}\frac{M^3-8a^2M}{8C}}
$$
is non-negative for the range of rotation parameter $|a|\leq3\sqrt{3}
M/16$ (see Fig.~\ref{region}), the equality indicates the extremal value
of the rotation parameter. It is $r_{+,-}$ that mark outer and inner horizons
of the hole.
The static limit is defined where the time-translation Killing
vector $\xi_{(t)}^\alpha$ becomes null (i.e. $g_{00}= \Delta - a^2 \sin^2\theta = 0$).
The region bounded by the outer horizon and the static limit defines the ergosphere (See Fig. 1), the extent of which increases with the rotation $a$ of the hole.

\section{Geodesics and circular orbits}\label{second}

In order to study particle motion around six dimensional pure GB black hole we first write the Hamilton-Jacobi equation,

\begin{equation}
\label{1}
\frac{\partial S}{\partial
\sigma}=-\frac{1}{2}g^{\mu\nu}\frac{\partial S}{\partial
x^{\mu}}\frac{\partial S}{\partial x^{\nu}} \ ,
\end{equation}
for Hamiltonian  \cite{frolov08,Frolov03,Papnoi14}:
\begin{equation}
\label{01}
S = \frac{1}{2}m^2 \sigma - {\cal E} t + {\cal L} \phi +
S_{r}(r)+S_{\theta}(\theta)+ {\cal W} \chi + T_{\Psi}(\Psi) \ ,
\end{equation}
where $m$ is mass, ${\cal E}$ and ${\cal L}$ are
conserved energy and angular momentum of the particle.

{The issue of the separability of the Hamilton-Jacobi equation in higher dimensional spacetime has been widely studied in the literature \cite{frolov08,Frolov03,Papnoi14}. Particularly, the authors of the paper~\cite{frolov08,Frolov03} have shown that the spacetime metric (\ref{metric}) is Petrov type D. }

\begin{figure*}[t!]

\includegraphics[width=0.32\linewidth]{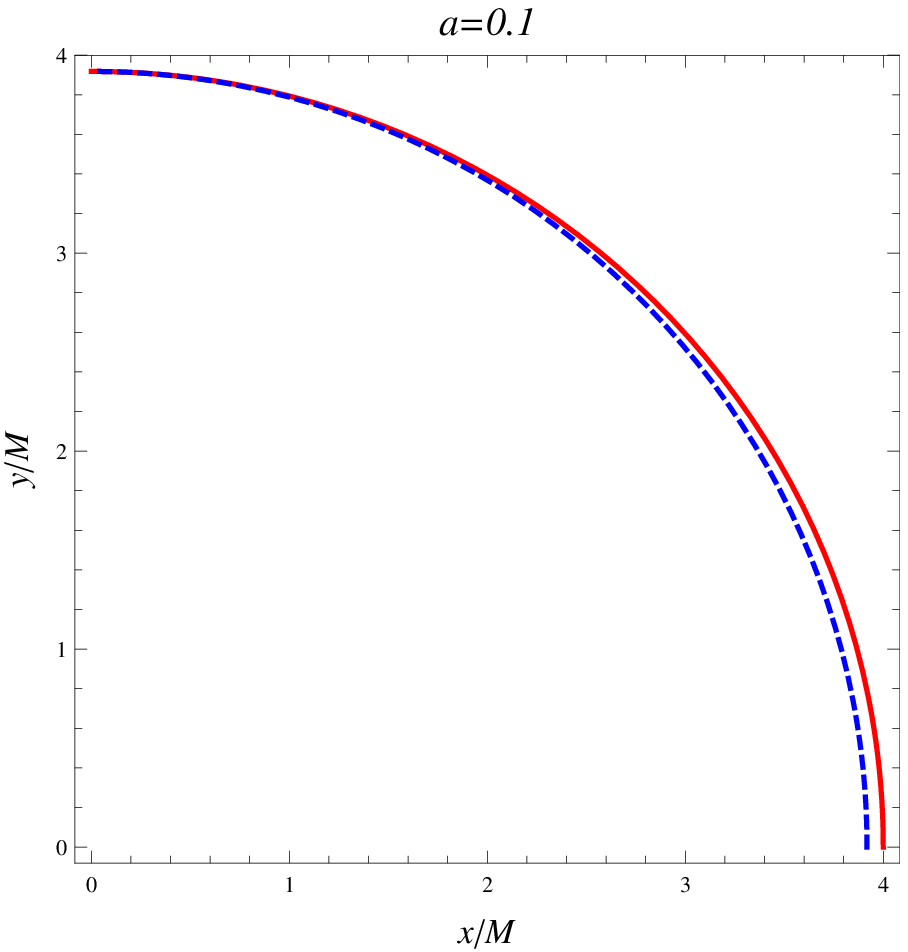}
\includegraphics[width=0.32\linewidth]{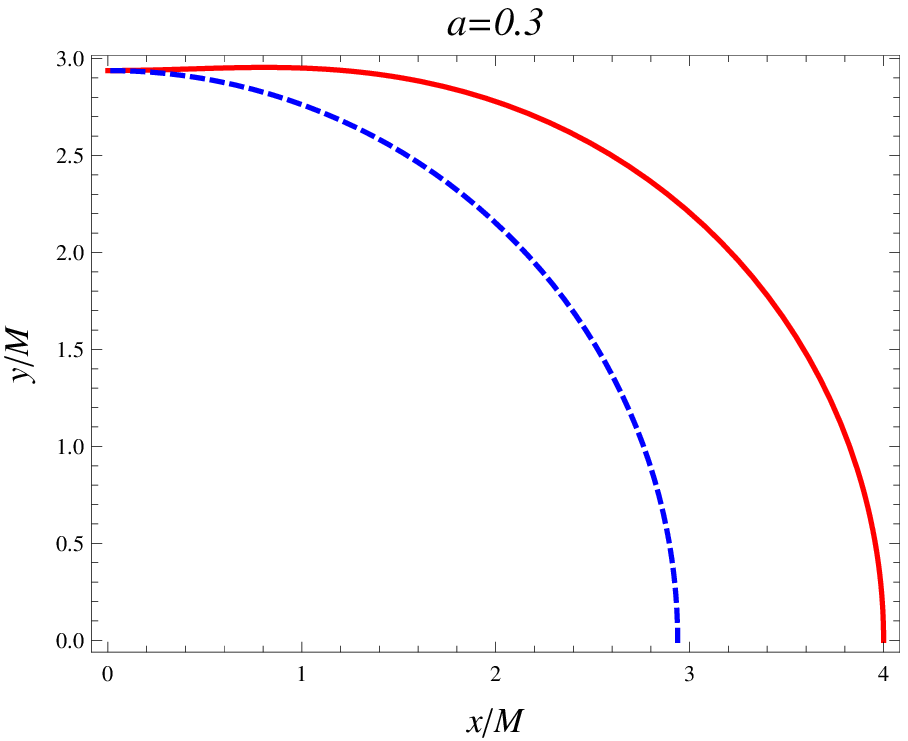}
\includegraphics[width=0.32\linewidth]{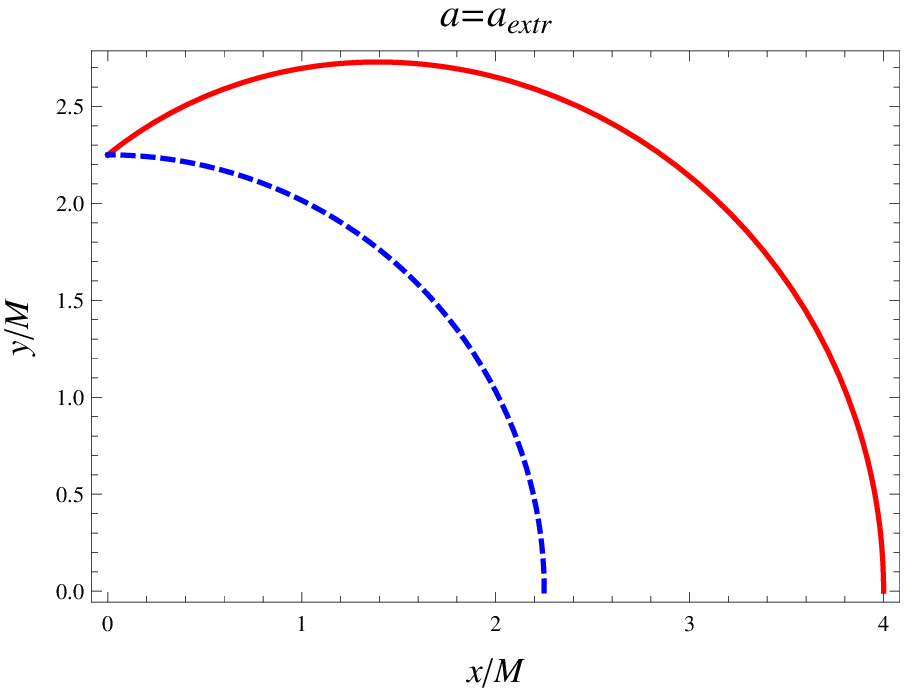}

\caption{The ergosphere for the different values of the spin parameter $a$:
$a/M=0.1$ (left panel),
 $a/M=0.3$ (middle panel), and  $a/M=a_{extr}/M=3\sqrt{3}/16$ (right panel).
 Red lines
indicate static limit while dashed blue lines indicate horizon.
 \label{ergs}}
\end{figure*}

\subsection{Null circular orbits}

For null geodesics (when m = 0), one can get the  equations of
motion from the Hamilton-Jacobi equation~(\ref{1}) as
\begin{eqnarray}
\Sigma\frac{dt}{d\sigma}&=&a({\cal L}-a {\cal E} \sin^2\theta)
+\frac{r^2+a^2}{\Delta} \Big[(r^2+a^2){\cal E}-a {\cal L}\Big] \ ,  \label{2}\\
\Sigma \frac{d\phi}{d\sigma}&=&(\frac{{\cal L}}{\sin^2\theta}-a {\cal E} )  
+\frac{a}{\Delta} \Big[(r^2+a^2){\cal E}-a {\cal L}\Big] \ ,\label{3}\\
\cos^2 \theta\frac{d\chi}{d\sigma}&=&\frac{ \csc^2 \psi }{r^2} {\cal W} \ , \label{4}\\
\Sigma \frac{dr}{d\sigma}&=&\sqrt{\mathcal{R}} \ ,
\label{5}\\
\Sigma \frac{d\theta}{d\sigma}&=&\sqrt{\Theta} \ , \label{6}
\\
\cos^2\theta \frac{d\psi}{d\sigma}&=&\frac{{\cal W} }{r^2 \sin\psi} \ , \label{7}
\end{eqnarray}
where new functions
$\mathcal{R}(r)$ and $\Theta(\theta)$
\begin{eqnarray}
\mathcal{R}&=&\left[(r^2+a^2){\cal E} -a {\cal L}
\right]^2-\Delta[\mathcal{K}+({\cal L} - a
{\cal E})^2] \ ,
\\
\Theta &=&\mathcal{K}-\frac{1}{\sin^2\theta} [a {\cal E}
\sin^2\theta-{\cal L} ]^{2} \
\end{eqnarray}\label{8}
are introduced.


The conserved quantity
$\cal W$
exists only when $\theta \neq \pi/2$ and is similar to
angular momentum. So called Carter constant $ \mathcal{K} $
characterizes
together with the quantities
$\cal E$, $\cal W$ and $\cal L$ the geodetic motion. The Carter
constant is not related to any isometry of the space-time unlike
the conserved quantities
$\cal E$, $\cal W$ and $\cal L$.

By defining $\xi={\cal L}/{\cal E}$, $\eta=\mathcal{K}/{\cal
E}^{2}$ and $ \zeta={\cal W}/{\cal E}$ and using Eq. (\ref{5}) we get for the circular
orbits characterized by
$\mathcal{R}(r)=0$ and $d\mathcal{R}(r)/dr=0$,
\begin{eqnarray}
\xi(r,a)&=&\frac{a^2(2r+3r^{1/2})+r^2(2r-5r^{1/2})}{a(3r^{1/2}-2r)} \ ,
\label{eqxi}
\\
\eta(r,a)&=&\frac{8a^2 r^{7/2} - r^4(2r-5 r^{1/2}
)^{2}}{a^2(2r-3r^{1/2})^2} \ . \ \ \ \ \label{eqeta}
\end{eqnarray}

%
%
\subsection{Timelike circular orbits}

\begin{figure*}[t!]

 \includegraphics[width=0.45\linewidth]{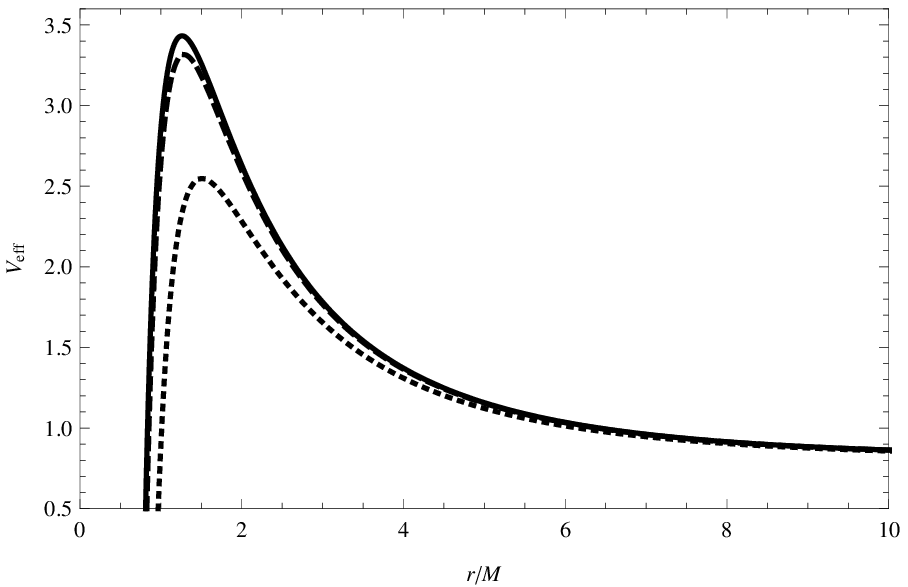}  \includegraphics[width=0.45\linewidth]{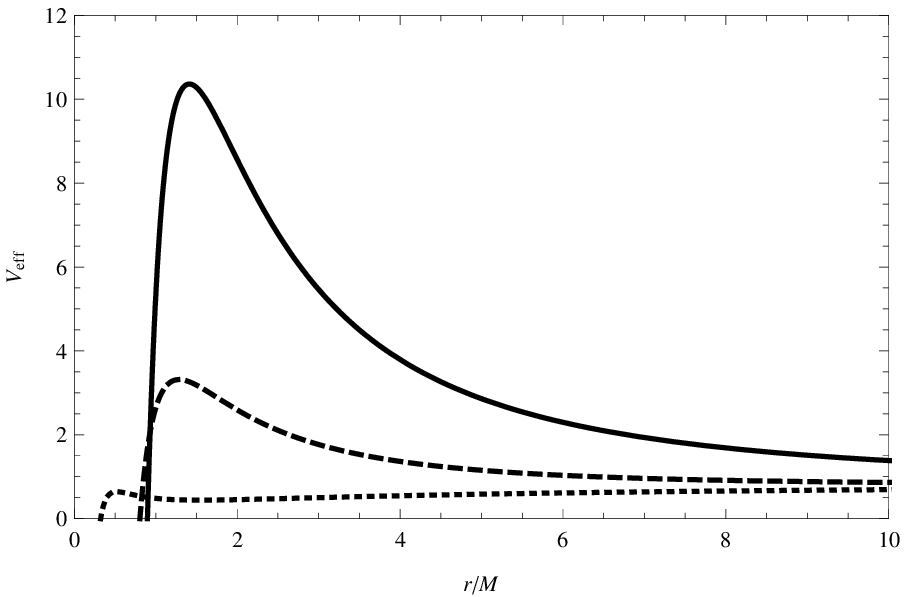}

\caption{The radial dependence of effective potential for the
different values of angular momentum and rotation parameter. The left plot corresponds to
the case when $a$ varies and the graphs are plotted for the different values of rotation parameter:  $a/M=0.1$ (dotted line),
 $a/M=0.3$ (dashed line), and  $a/M=a_{extr}/M=3\sqrt{3}/16$ (solid line).
 The right plot  corresponds the case to
the case when $L$ varies and the graphs are plotted for the different values of the angular momentum of the particle:
$L/Mm= 1$ (dotted line), $L/Mm= 5$ (dashed line), and $L/Mm= 10$ (solid line) \label{effpot}}
\end{figure*}

Consider the equation of motion of test particle with nonzero rest mass at the equatorial
plane ($\theta=\pi/2\ ,\ \dot{\theta}=0$). The equations of motion take the following form
\begin{eqnarray}
\label{tuch}
&&m \frac{dt}{d\sigma}=-\frac{{\cal E} r^{2}-a {\cal L} \sqrt{M/r} + a^2 {\cal }E \left(\sqrt{M/r} + 1\right) }
{a^2 - M^{1/2} r^{3/2} + r^2} \ , \\
&&m \frac{d\phi}{d\sigma}=\frac{{\cal L} + (a {\cal E} - {\cal L}) M^{1/2} {r}^{-1/2}}{a^2 - M^{1/2} r^{3/2} + r^2}\ , \\
&& m^2\left(\frac{dr}{d\sigma}\right)^2={\cal E}^2-V_{\rm eff}\ ,
\label{ruch}
 \end{eqnarray}
where
\begin{eqnarray}
V_{\rm eff}&=&m^2\left(1-\sqrt{\frac{M}{r}}\right) - \frac{(
{\cal L}-a {\cal E} )^2 M^{1/2}}{r^{5/2}}\nonumber\\ &&
+ \frac{
 {\cal L}^2 - a^2 ({\cal E}^2 - m^2)}{r^2}
\end{eqnarray}
is the effective potential for radial motion. Note that for the orbits in the equatorial plane the new
conserved quantity $\cal W$ does not appear in the equations of
motion. {In Fig.~\ref{effpot} the radial dependence of the
effective potential of radial motion in equatorial plane has been
shown for the fixed specific value of the rotation parameter
$a=M$. The increase of the momentum of the particle leads to the
increase of the peak of the potential: initially infalling test
particles become bounded or escaped with the increase of the
momenta. }

The conditions of occurrence of circular orbits are
$$
\frac{dr}{d\sigma}=0\,,\ V_{\rm eff}'(r)=0\ .
$$

From these equations, it follows that energy ${\cal E}$ and
angular momentum ${\cal L}$ of a circular orbit of radius $r_{c}$
are given by
%
\begin{eqnarray}
{\cal E}^2&=&\bigg[a^2(12 \sqrt{r}-11)r^2 -4 (\sqrt{r}-1)^2(4\sqrt{r}-5) r^{7/2} \nonumber\\ && \pm 4  a r^{5/4} (a^2 - r^{3/2} + r^2)\bigg]\nonumber\\ && \times \bigg[{ 16 a^2 r^{5/2}-(5 - 4 \sqrt{r})^2 r^4  }\bigg] \ ,\\
{\cal L}^2&=&\bigg[5 r^6- a^4 (11 + 4 \sqrt{r}) r^2  - 4 r^{13/2} \nonumber\\&& +2 a^2 (10 - 11 \sqrt{r} - 4 r) r^{7/2} \nonumber \\ && \pm (a^2- r^{3/2}+r^2)(20 r^2 +4a^2)  \bigg]\nonumber\\&&\times \bigg[{16 a^2 r^{5/2}-(5 - 4 \sqrt{r})^2 r^4 }\bigg]^{-1}\ ,
\end{eqnarray}
where $+$ and $-$ signs correspond to the co-rotating and
counter-rotating particles.

\begin{figure*}[t!]
 \includegraphics[width=0.45\linewidth]{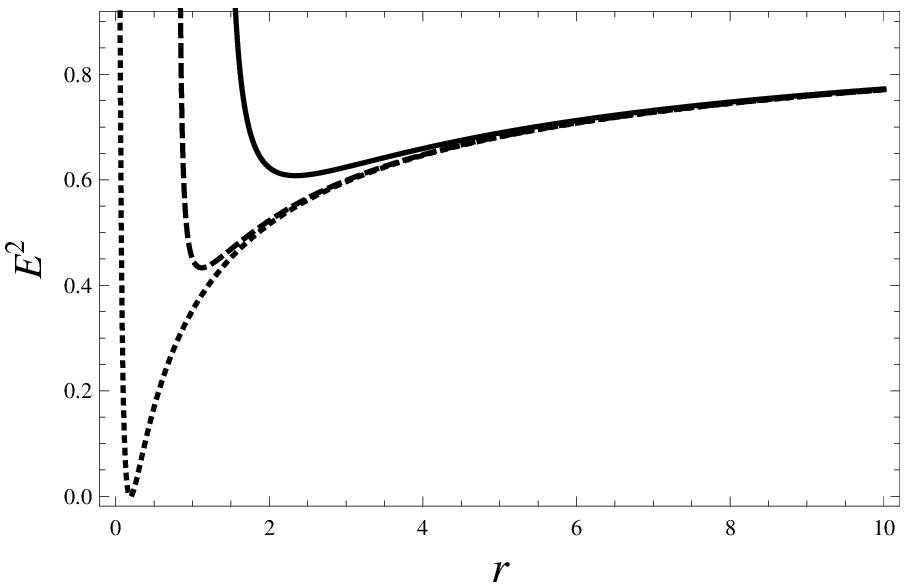}
 \includegraphics[width=0.45\linewidth]{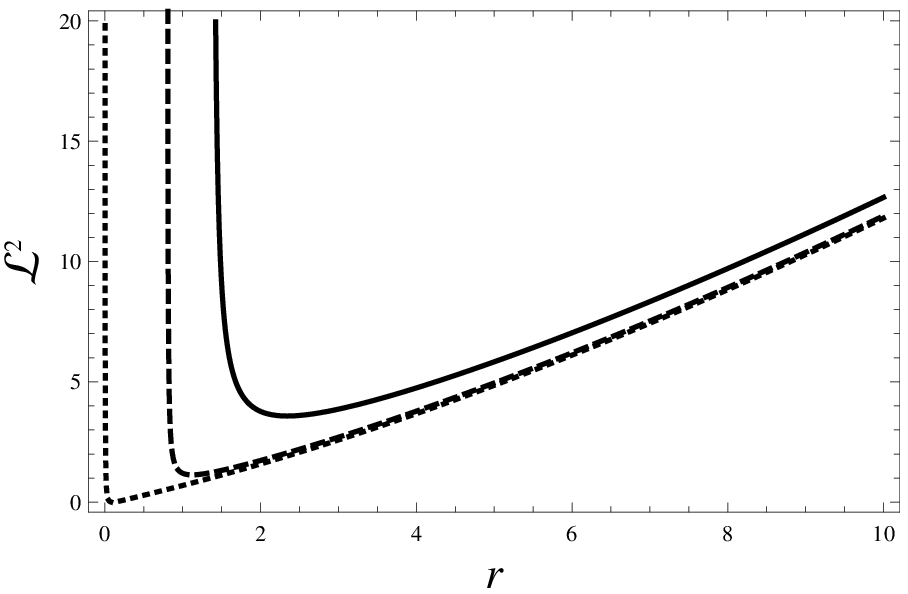}

 \includegraphics[width=0.45\linewidth]{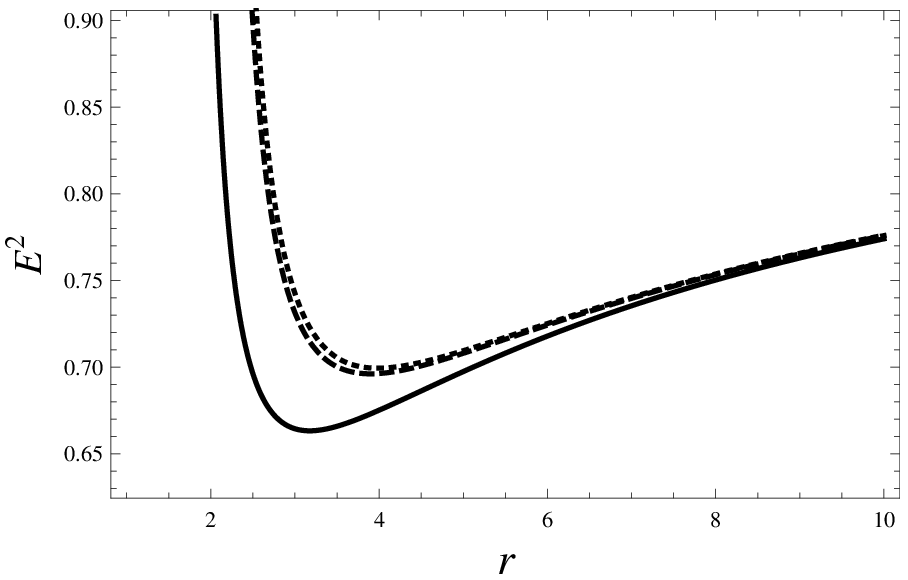}
 \includegraphics[width=0.45\linewidth]{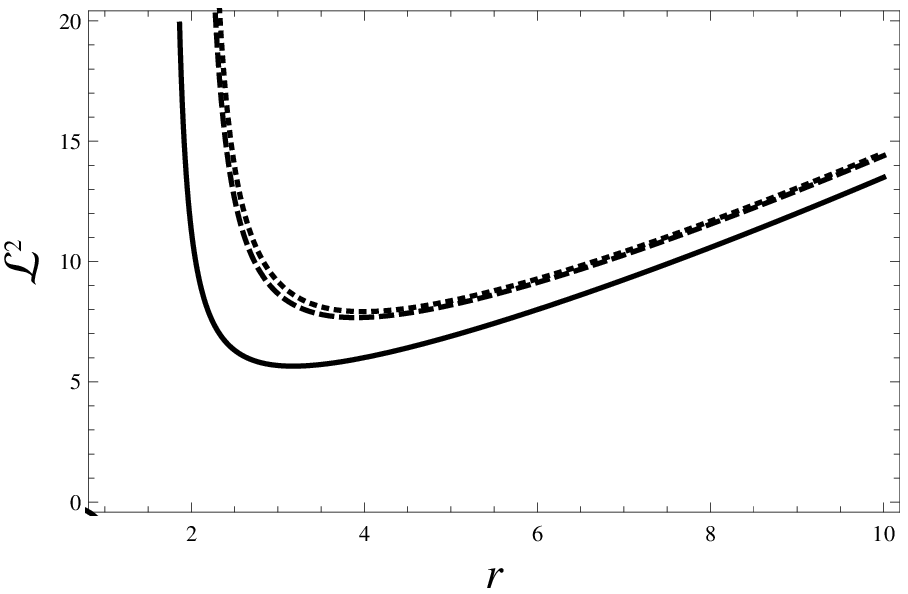}
\caption{The radial dependence of energy (left) and angular
momentum (right) squared of counter-rotating (upper plots) and
co-rotating (lower plots) particle at circular orbits  for
the different values of parameter $a= 0.1$ (solid line), $0.3$ (dashed
line), and $3\sqrt{3}/16$ (dotted line). \label{enmom}}
\end{figure*}

In the Fig.~\ref{enmom} we have shown energy and angular
momentum of the co and counter rotating orbits in the equatorial
plane for different values of rotation parameter $a$. One can easily see the shift in
location of the minimal circular orbit (MC) marking the existence limit given by the photon circular
orbit and the innermost stable circular orbits (ISCO). The MC and ISCO come closer to the hole with
increase in $a$ for co-rotating orbits while opposite happens for counter rotating ones.

\begin{figure}[t!]
\begin{center}
\includegraphics[width=0.8\linewidth]{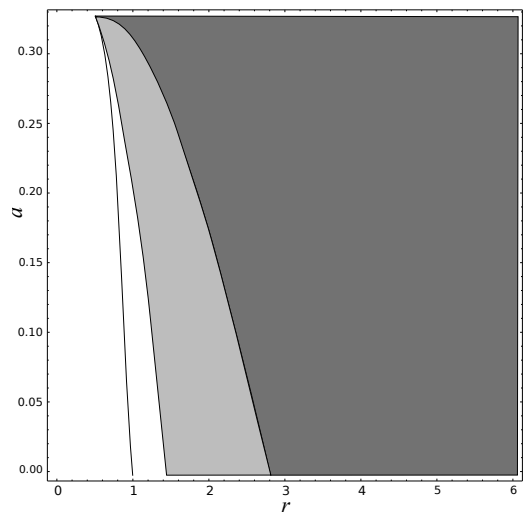}
\end{center}

\caption{The regions of stable (dark grey) and unstable (light
grey) circular orbits.  The black curve indicates the border of
event horizon of 6-D Gauss-Bonnet black hole. \label{region}}
\end{figure}

The vanishing of denominator in the expressions for ${\cal E}$ and ${\cal L}$ marks
the location of photon circular orbit or MC while for ISCO we have $dr/d\sigma=V_{\rm eff}'(r)=0$
and $V_{\rm eff}''(r)\geq 0$.
In Fig.~\ref{region} we have shown three regions as dark, light grey and white marking the boundaries of
stable, unstable and no circular orbits. The inner boundary of light grey is defined by photon circular
orbit and the white region bounded between it and the horizon is the one where no circular orbits can
exist. This is the region between 3M and 2M for the Schwarzschild black hole. As expected these regions are
quite similar to that of the 4-dimensional Kerr black hole.

At this point it may be mentioned that bound orbits cannot exist for Einstein gravity in dimensions
$\geq4$ [7] in general and in particular their non-existence is shown for $6$-dimensional rotating black hole in
Ref.~\cite{hackmann08}. For pure Lovelock gravity they do always exist in all even dimensions, $d=2N+2$ [7].

\begin{figure*}[t!]
\begin{center}
\includegraphics[width=0.26\linewidth]{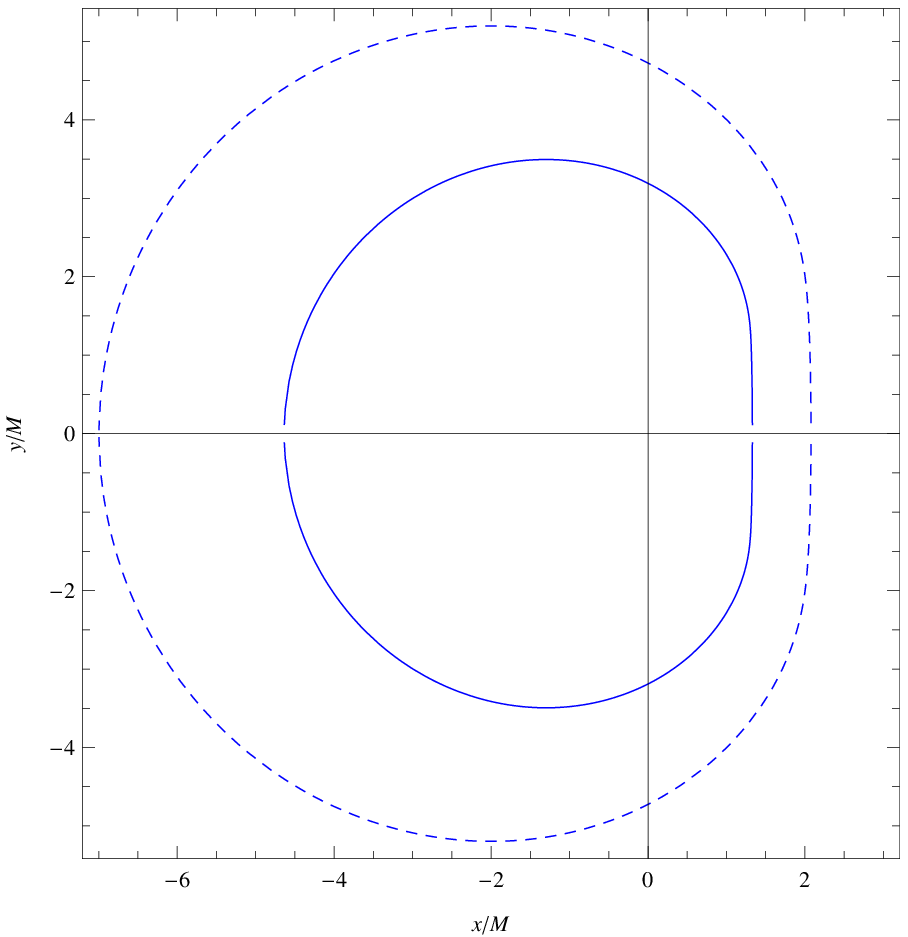}
\includegraphics[width=0.31\linewidth]{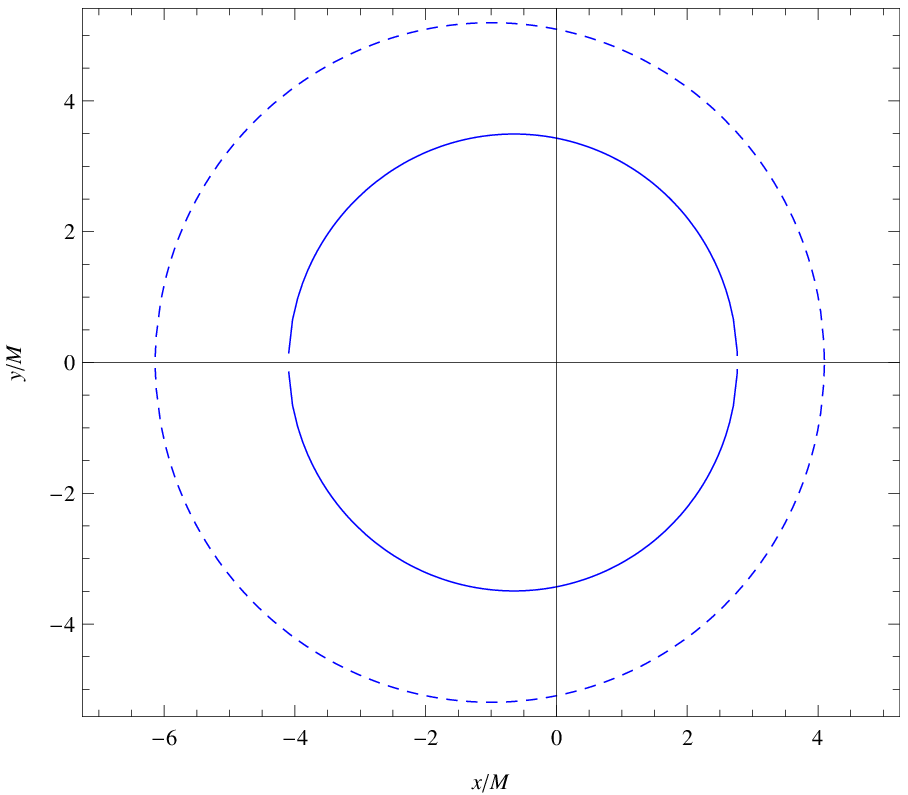}
\includegraphics[width=0.31\linewidth]{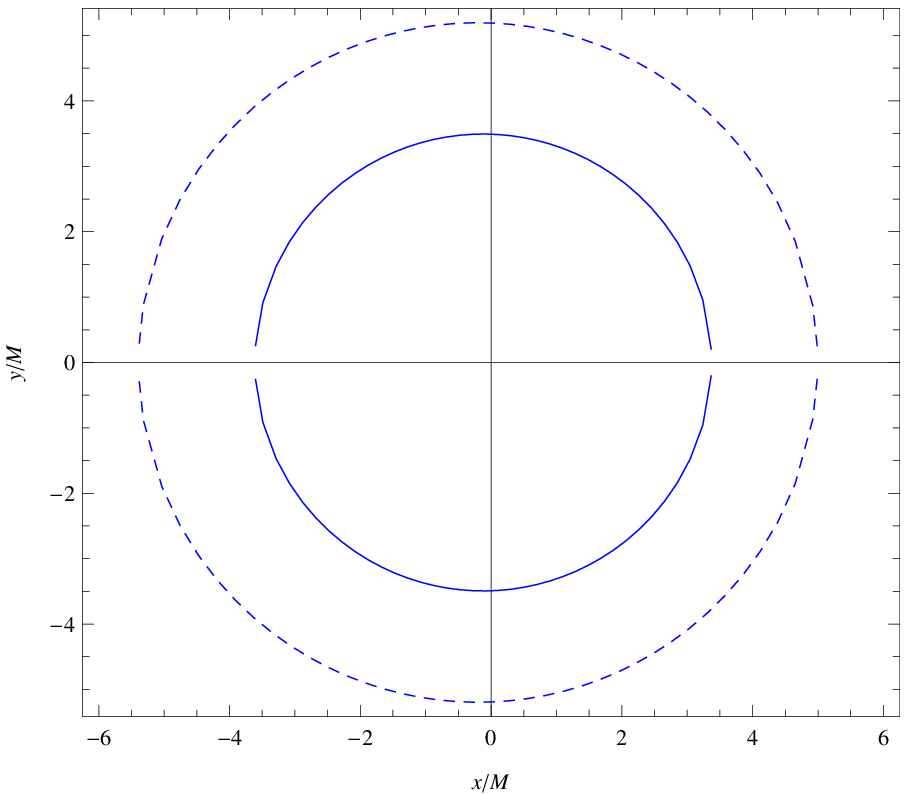}

\end{center}
\caption{ Shadow of the rotating GB black holes   when the
inclination angle $\theta = \pi/2$. For the comparison we present
the shadow of rotating Kerr black hole (dashed lines). From
left to right, the rotation parameter scans as  $a=0,\
\ a_{\rm ext}/2, \ \ {\rm and }\ \ a_{\rm ext} $. Note that for the pure  GB 6-D black hole $a_{\rm ext}=3\sqrt{3}/16$ as against   $a_{\rm ext}=1$ for the Kerr black hole. \label{shadow1}}
\end{figure*}

\begin{figure*}[t!]
\begin{center}
\includegraphics[width=0.4\linewidth]{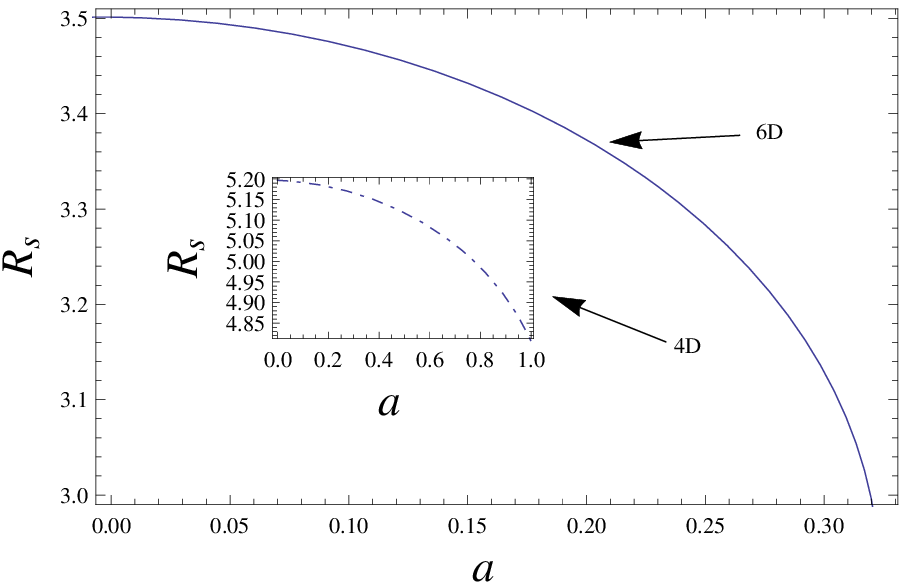}
\includegraphics[width=0.4\linewidth]{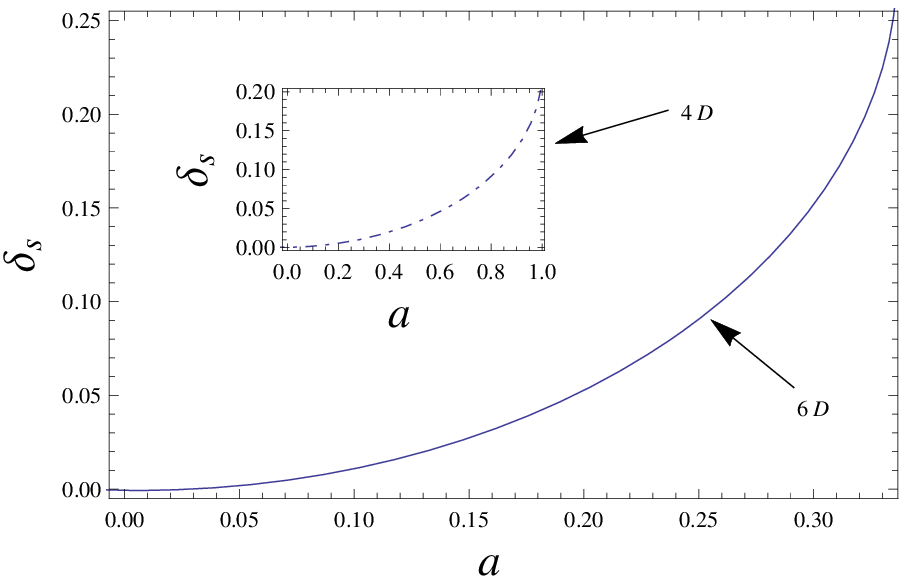}
\end{center}
\caption{  Observables $R_{s}$ and $\delta_{s}$ as functions of
the rotation parameter, corresponding to the shadow of  black hole
situated at the origin of coordinates with inclination angle
$\theta=\pi/2$ and dimensions $d$: $D=6$ (solid line) and $d=4$
(dot-dashed line)~\cite{Hioki09,Zaharov}. \label{shadow4}}
\end{figure*}

\begin{figure*}[t!]
\begin{center}
\includegraphics[width=0.4\linewidth]{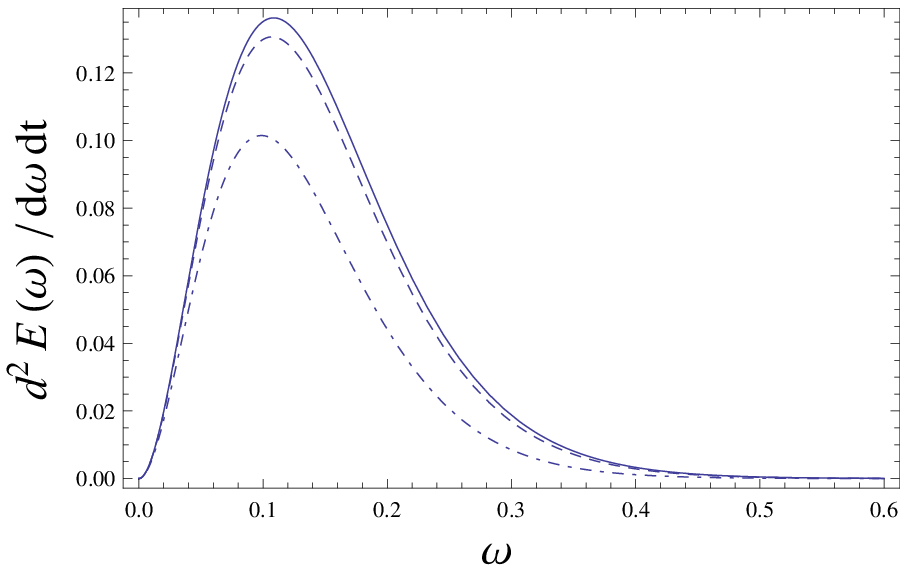}
\includegraphics[width=0.4\linewidth]{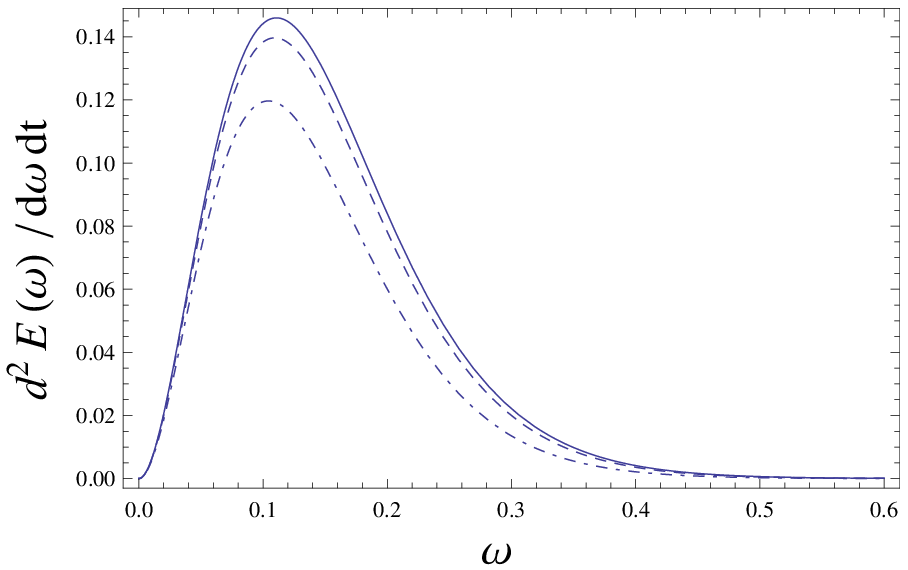}
\end{center}
\caption{Dependence of the energy emission from the frequency for
the different values of spin parameter $a$: $a/M=0.1$ (solid
line), $a/M=0.3$ (dashed line), $a/M=a_{extr}/M=3\sqrt{3}/16$ (dot-dashed line). Left
panel is  for black hole in the Gauss-Bonnet and right panel is for 4D
black hole in the Kerr space-time \cite{wei}. \label{energy}}
\end{figure*}

\section{Black hole shadow \label{third}}

In this section we study the optical properties of  black hole  in Gauss-Bonnet
gravity. If the bright source is located behind black hole then
a distant observer  is able to detect only photons scattered away
from the black hole, and those captured by the event horizon
form a dark spot. This dark region which could be detected and extracted from the luminous background is
traditionally called the
black hole shadow or silhouette.
In practice, the distant observer
at infinity could see a projection of it at the flat plane passing through the black hole
and normal to the line connecting it with the observer
(the line of sight). The Cartesian coordinates at this plane, which
are usually denoted by  $\alpha$	 and $\beta$ and called celestial coordinates,
 give the apparent position of the shadow image. The celestial coordinates are connected with the geodesic
equations of photons around black hole as~\cite{Vazquez}
\begin{eqnarray}  \label{alpha}
\alpha&=&\lim_{r_{0}\rightarrow \infty}\left(
-r_{0}^{2}\sin\theta_{0}\frac{d\phi}{dr}
\right) \ ,
\\
\label{beta1}
\beta&=&\lim_{r_{0}\rightarrow \infty}r_{0}^{2}\frac{d\theta}{dr} \ ,
\end{eqnarray}
with $r_{0}\rightarrow\infty$, $\theta_{0}$ is the inclination
angle between the line of sight of the far observer and  the
 axis of rotation of the black hole \cite{Amarilla12}, also see \cite{Atamurotov13b}.

{Note here that a silhouette of the black hole is observed in 3-D
space and here we would like to check the influence of the extra dimensions to the shape of the
black hole shadow.}

With the help of the expressions for the impact parameters derived in the Sec.
\ref{second} and the equations of motion obtained from the Hamilton-Jacobi
equation (\ref{1}), one can get $d\phi/dr$ and $d\theta/dr$, and
insert them into the equations (\ref{alpha}) and (\ref{beta1}) in
order to get the explicit expressions for  the celestial coordinates as
\begin{eqnarray}  \label{alphapsi1}
\alpha&=&-\xi\csc\theta_{0}
\ ,
\\  \label{beta2}
\beta&=&\pm \sqrt{\eta + a^2 \cos ^{2}\theta_{0}-\xi \cot
^{2}\theta_{0}} \ .
\end{eqnarray}

{We will concentrate here on the special case when the
inclination angle $\theta_0=\pi/2$ is similar to that for four dimensional Kerr space-time~(see
e.g.~\cite{Stuchlik12a,Amarilla10,Amarilla13,Hioki09}). Then for the pure GB 6-D rotating black hole we have
\begin{eqnarray}  \label{alphapsi2}
\alpha&=&-\xi \ ,
\\  \label{beta3}
\beta&=&\pm \sqrt{\eta} \ .
\end{eqnarray}

To get boundary of the black hole shadow one can plot dependence of the coordinate 
$\beta$ from the coordinate $\alpha$, see e.g.~\cite{Atamurotov13b}. In Fig.
\ref{shadow1}, we compare shadow of six dimensional black hole in
Gauss-Bonnet gravity with one of four dimensional Kerr black hole,
which are shown for the different values of the
rotation parameter $a$. The contours of the shadows of the
Gauss-Bonnet black hole for the spin parameters $a=0$, $a=a_{ext}/2$,
and $a=a_{ext}$ are shown in Fig. \ref{shadow1}. One
can easily see, photon sphere is decreased with the increase of
spin of the black hole in Gauss-Bonnet gravity. This behavior is
exactly the same as in the Kerr space-time.

The observable parameters as distortion parameter $\delta_{s}$ and
radius of the shadow $R_s$ can be computed numerically using
either the expressions (\ref{alphapsi2}) and  (\ref{beta3}) or
Fig. \ref{shadow1}. Distortion parameter $\delta_{s}=\Delta x/
R_s$ \cite{Atamurotov13b,Hioki09}, where $\Delta x$ is deviation parameter which
is distance between edge point of full circle and edge point  of
shadow \cite{Atamurotov13b}.  Consequently if rotation parameter is equal to
zero $a=0$ then $\Delta x$ must vanish. On the other hand, if we
consider rotating black hole, $\Delta x$ is nonzero and
consequently $\delta_{s}$ depends on spin of black hole.
In Fig.~\ref{shadow4},  the observables $R_{s}$ and $\delta_{s}$
as functions of the rotation parameter of the black hole are shown
when the inclination angle $\theta_{0}=\pi/2$.
From these plots one may conclude that with the increase  of spin
parameter $a$ of the black hole in Gauss-Bonnet gravity shape of
shadow is decreasing which is similar to the Kerr black hole case.
The increase of $\delta_s$ with the increase of rotation
parameter $a$ corresponds to deviation of the shape of shadow from
circle.

\section{\label{enextract}Energetics \label{four}}

\subsection{Emission energy of 6D rotating black hole}

As the next step we plan to calculate energy emission from rotating black
hole in higher dimensional Gauss-Bonnet gravity as~\cite{wei}
\begin{equation}
\frac{d^2 E(\omega)}{d\omega dt}=\frac{2\pi^3
R^2_{s}}{e^{\omega/T}-1} \omega^3 \ ,
\end{equation}
where $\omega$ is the frequency of the emission,\\
$T=\left({r_{+}^2-3a^2}\right)/\left({8 \pi r_{+} (r_{+}^2+a^2)}\right)$ is the Hawking
temperature for the Gauss-Bonnet black hole (for comparison,  
$T=\left({r_{+}^2-a^2}\right)/\left({4 \pi r_{+} (r_{+}^2+a^2)}\right)$ is the Hawking
temperature of the Kerr black hole \cite{wei}), 
which can be computed from this expression $T=k/ 2\pi$, $k$ is surface gravity.
$R_{s}$ is radius of shadow which is shown in Fig. \ref{shadow4}
for the second order Lovelock space-time \cite{wei}.

The comparison of the energy emission of rotating black hole in
Gauss-Bonnet and Kerr space-time for the different values of spin
parameters $a$: $a=0.1$ (solid line), $a=0.2$ (dashed line),
$a=0.3$ (dot-dashed line) is represented in Fig. \ref{energy}. 
The rate of energy emission decreases as the rotation parameter increases. The emission
is more intense for the Kerr black hole as compared to GB one.

\subsection{Particle acceleration through BSW effect}

Here first we define the energy $E_{\rm cm}$ in the center of mass
of system of two colliding particles with energy  at infinity
$E_{1}$ and $E_{2}$ in the gravitational field described by
spacetime metric (\ref{metric}) as
\begin{equation}
E_{\rm cm}^2= p_{\rm (tot)}^{\alpha}p_{{\rm (tot) }\alpha}\ ,
\end{equation}
where $p_{\rm (tot)}^{\alpha}=p_{\rm (1)}^{\alpha}+p_{\rm
(2)}^{\alpha}$ is the total momenta of particles 1 and 2 with the
mass  $m_{1}$, $m_{2}$, respectively. We assume that two particles
with equal mass ($m_{1}=m_{2}=m_{0}$) have the energy at infinity
$E_{1}=E_{2}\simeq 1$, and consequently
\begin{eqnarray}
E_{\rm cm}= m_{0} \sqrt{2} \sqrt{1- g_{\alpha\beta}
v_{(1)}^{\alpha} v_{(2)}^{\beta}}\ .
\end{eqnarray}

Now using the equations (\ref{tuch})--(\ref{ruch}) we derive
expression for the center-of-mass energy of particles in collision
 in the vicinity of the
Gauss-Bonnet   black hole as
%
\begin{eqnarray}
\frac{E_{\rm c.m.}^2}{2m^2} & = & \frac{1}{\sqrt{r} (a^2 - r^{3/2} + r^2)}\nonumber\\&&\times \bigg(a^2 (1 + 2\sqrt{r}) -
 r^2-a (l_1 + l_2)\nonumber\\&& - l_1 l_2 ( \sqrt{r}-1)  + 2 r^{5/2} -
 r^{5/2} \nonumber\\ && \times\sqrt{a^2 - 2 a l_1 - l_1^2 (\sqrt{r}-1) + r^2}
    \nonumber\\ && \times\sqrt{a^2 - 2 a l_2 - l_2^2 ( \sqrt{r}-1) +  r^2}\bigg)\ ,
\end{eqnarray}
%
where we put $M=1$ and $l_1={L_1}/{m_1}$,
$l_2={L_2}/{m_2}$.

For the extremal rotating   Gauss-Bonnet black hole when
$a=3\sqrt{3}/16$ the center of mass energy at the horizon has the
following limit
\begin{eqnarray}
\frac{E_{\rm c.m.}^2}{2m^2}(r\rightarrow
r_{+})=\sqrt{\left(\frac{3\sqrt{3}-4l_1}{3\sqrt{3}-4l_2}\right)^3+\left(\frac{3\sqrt{3}-4l_2}{3\sqrt{3}-4l_1}\right)^3}\
.
\end{eqnarray}

Now we study the maximal energy which  can be extracted through
the BSW process~\cite{Banados09} discussed in the Introduction from
rotating Gauss-Bonnet black hole. For this purpose, first we need
to study the energy of the test particle moving on the
innermost stable circular orbit. Then we define the coefficient of
total amount of released energy of the test particle on shifting from
its stable circular orbit with the radius  $r_c$ to the
ISCO with the radius $r_{\rm ISCO}$ . The energy release
efficiency coefficient can be given as
\begin{eqnarray}
\eta=100\times \frac{E(r_{c})-E(r_{_{ISCO}})}{E(r_{c})} \ .
\end{eqnarray}
%
The radial dependence of the efficiency coefficient $\eta$  for the
different values of the rotation parameter $a$ is shown in
Fig.~\ref{efficiency}. The maximal energy extraction is for extremal black hole
for which $ r_{_{ISCO}}=r_{h}$ and so we obtain the maximal limit as $\eta\simeq55.28 \%$.

\begin{figure}[t!]
 \includegraphics[width=0.9\linewidth]{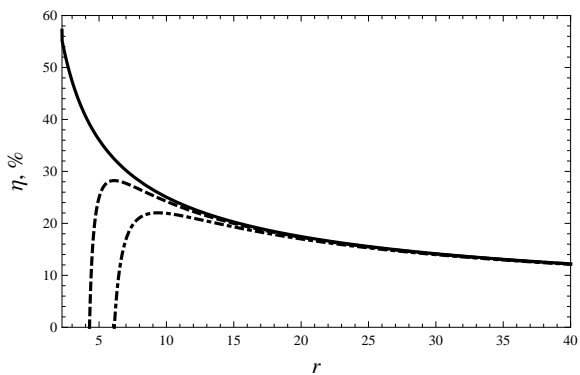}

\caption{The radial dependence of energy extraction efficiency
for the different values of the rotational parameter: $a=0.1$
(dot-dashed line), $a=.3$ (dashed line), and $a=3\sqrt{3}/16$ (solid
line). \label{efficiency}}
\end{figure}

\subsection{Penrose process}

The existence of an ergosphere around the   rotating black hole,
where negative energy states for the particles moving along the
timelike or null trajectory present, gives us opportunity to
consider the energy extraction from the rotating black hole
through Penrose process.  Assume a small particle $A$ falls down
into the ergosphere of the black hole from far outside. In the
vicinity of the event horizon it splits into two fragments, $B$ and
$C$. If particle $B$ with the negative energy with respect
to infinity falls into the central black hole then the emergent particle $C$
has the energy exceeding the energy  of the incident particle $A$.

\begin{eqnarray}
\label{EnergyF}
\alpha E^{2}-2\beta E+\gamma+\frac{\Sigma}{\Delta}(p^{r})^{2}+
\Sigma(p^{\theta})^{2}&&\nonumber\\+r^2\cos^2\theta (p^{\psi})^2+ m^{2}&=&0,\nonumber\\
\end{eqnarray}

where we have used the following notations
\begin{eqnarray}
\alpha&=&\Sigma\left(a^4+2a^2 r^2+r^4-a^2\Delta \sin^2\theta\right)\Gamma^{-1},\\
 \beta&=&-2M^{1/2}ar^{3/2}L\Sigma \Gamma^{-1} , \\
\gamma&=&\frac{L^2 \Sigma (a^2 \sin^2\theta - \Delta)}{\Gamma \sin^\theta}+\frac{W^2}{r^2\cos^2\theta \sin^2\psi} ,\\
\Gamma&=& a^2[(a^2+r^2)^2+\Delta^2-M r^3]\sin^2\theta\nonumber\\&&-\Delta (a^2+r^2)^2-a^4\Delta \sin^4\theta\ .
\end{eqnarray}

As the particle falls inside the event horizon the change of  the
mass of central rotating black hole defined as $\delta M=E$. In
principle one can increase the mass of the black hole increasing
the number of the infalling test particles with positive energy.
The  minimum value of the central black hole mass  $\delta M$ is
achieved for condition when
$m=0$, $p^{\theta}=0$, $p^{\psi}=0$, and $p^{r}=0$.
Then one can get the expression for the minimum energy:
\begin{eqnarray}
E_{\rm min}=\Omega(r_{+}) {\cal L} \ ,
\end{eqnarray}
where we have used the notation
$$
\Omega(r_{+})=-\left.\frac{g_{03}}{g_{33}}\right|_{r=r_+}=\frac{2aM^{1/2}  r_{+}^{3/2}}{a^2+r_{+}^2}\
$$
using the values of  the mass of the particle and momenta for the
minimum condition.

{Next we discuss the energy extraction efficiency from
Gauss-Bonnet black hole through  Penrose process. As in the case
of BSW process we  introduce the coefficient of efficiency of the
Penrose process as
\begin{eqnarray}
\eta_P=\frac{E_{C}-E_{B}}{E_{A}}\times 100\ ,
\end{eqnarray}
where $E_{A}$ is the energy of the incident particle and $E_{C}$  is that of
the emergent outgoing one. Using the energy conservation law
for the particles $A$, $B$ and $C$ one can find the maximal value
of $\eta_P$ in the form
\begin{eqnarray}
\eta_{P (max)}=\left[(\sqrt{1+g_{tt}}+1)/2-1\right]\times 100\ .
\end{eqnarray}
Evaluating this expression near the event horizon of    extreme
rotating Gauss-Bonnet black hole one can find the maximal
efficiency of the value of $25.8 \%$. Note that the energy
extraction efficiency for the Penrose process in the case of
extreme rotating 4-D Kerr black hole was found to be $20.7
\%$~\cite{chandra98,Part}. }

\section{Discussion}
\label{remarks}

The elemental feature of pure Lovelock gravity is that
gravitational dynamics in all odd and even dimensions is similar.
That is why it is expected that physical processes and effects
around a $6$-dimensional rotating pure GB black hole would be
similar to that of $4$-dimensional Kerr black hole. It should
however be admitted that the black hole metric we have considered
has all the desired features of a rotating black hole but it is
though not an exact solution of the pure Lovelock vacuum equation.
It does however satisfy the equation in the leading order which is
computed as follows. Since metric goes as $r^{-1/2}$, Riemann tensor will
go as $r^{-5/2}$, and then GB Ricci tensor will go as $r^{-5}$. For the
black hole metric (2), GB Ricci tensor in fact falls off as $r^{-7}$, two
powers sharper. It could therefore be taken as a good model for
describing  a rotating pure GB black hole.

As gravitational potential is weaker than  Einstein gravity, its
effects are reflected as follows. The efficiency of Penrose
process decreases and it is reduced to   $7.74\%$ (it is equal to
is $29\%$ for the 4-D Kerr black hole)  while the opposite is
effect on particle acceleration efficiency which is increased to
$55.28\%$ (it is equal to $46\%$ for the 4-D Kerr black hole). The
center-of-mass energy rapidly grows
 for a collision of particles falling from infinity into
rotating   GB black hole in the case when the  circular orbits
shift arbitrarily close to the horizon.  The optical shadow of
black hole also decreases as lesser number of photons get captured
because of weakening of the field. All these results are in on the
expected lines on physical grounds, and hence they provide
strength to validity and viability of the spacetime metric used. In other
way,  this study could be looked upon probing the metric in
question for its in principle physical and astrophysical validity.

We had set out to study various physical properties of a pure GB rotating black hole
and show that they are indeed similar to the rotating black hole in the usual four dimensional
physical spacetime. All this is in line with the pure Lovelock gravity paradigm \cite{dadhich15}
in higher dimensions.

\section*{Acknowledgments}

{The~authors acknowledge the~project Supporting Integration with
the~International Theoretical and Observational Research Network
in Relativistic Astrophysics of Compact Objects,
CZ.1.07/2.3.00/20.0071, supported by Operational Programme
\emph{Education for Competitiveness} funded by Structural Funds of
the~European Union. One of the authors (ZS) acknowledges the
Albert Einstein Center for gravitation and astrophysics supported
by the~Czech Science Foundation No.~14-37086G.} Warm hospitality
that  has facilitated this work to A.A., B.A. and N.D. by Faculty
of Philosophy and Science, Silesian University in Opava (Czech
Republic) and  by the Goethe University, Frankfurt
am Main, Germany is thankfully acknowledged. This research is
supported in part by Projects No. F2-FA-F113, No. EF2-FA-0-12477,
and No. F2-FA-F029 of the UzAS and by the ICTP through the
OEA-PRJ-29 and the OEA-NET-76 projects and by the Volkswagen
Stiftung (Grant No. 86 866). A.A. and B.A. acknowledge the TWAS
associateship grant.


\bibliography{gravreferences}

\providecommand{\newblock}{}
\begin{thebibliography}{10}
\expandafter\ifx\csname url\endcsname\relax
  \def\url#1{{\tt #1}}\fi
\expandafter\ifx\csname urlprefix\endcsname\relax\def\urlprefix{URL }\fi
\providecommand{\eprint}[2][]{\url{#2}}

\bibitem{lovelock}
{Lovelock} D 1971 {\em J. Math. Phys.\/} {\bf 12} 498--501

\bibitem{dadhich04}
{Dadhich} N 2004 {\em ArXiv General Relativity and Quantum Cosmology
  e-prints\/} (\textit{Preprint} \eprint{gr-qc/0405115})

\bibitem{dadhich05}
{Dadhich} N 2005 {\em ArXiv High Energy Physics - Theory e-prints\/}
  (\textit{Preprint} \eprint{hep-th/0509126})

\bibitem{dadhich11}
{Dadhich} N 2011 {\em International Journal of Modern Physics D\/} {\bf 20}
  2739--2747 (\textit{Preprint} \eprint{1105.3396})

\bibitem{banados94}
{Ba{\~n}ados} M, {Teitelboim} C and {Zanelli} J 1994 {\em Physical Review
  Letters\/} {\bf 72} 957--960 (\textit{Preprint} \eprint{gr-qc/9309026})

\bibitem{dadhich12a}
{Dadhich} N 2012 {\em ArXiv e-prints\/} (\textit{Preprint} \eprint{1210.3022})

\bibitem{Dadhich12c}
{Dadhich} N, {Ghosh} S~G and {Jhingan} S 2012 {\em Physics Letters B\/} {\bf
  711} 196--198 (\textit{Preprint} \eprint{1202.4575})

\bibitem{Camanho15}
{Camanho} X~O and {Dadhich} N 2015 {\em ArXiv e-prints\/} (\textit{Preprint}
  \eprint{1503.02889})

\bibitem{Penrose69}
Penrose R 1969 {\em Riv. Nuovo Cimento\/} {\bf 1} 252

\bibitem{wagh85}
{Wagh} S~M, {Dhurandhar} S~V and {Dadhich} N 1985 {\em Astrophys J.\/} {\bf
  290} 12--14

\bibitem{wagh86erratum}
{Wagh} S~M, {Dhurandhar} S~V and {Dadhich} N 1986 {\em Astrophys J.\/} {\bf
  301} 1018

\bibitem{wagh85a}
{Wagh} S~M, {Dhurandhar} S~V and {Dadhich} N 1985 {\em in Quasars, eds G.
  Swarup and V. Kapahi (Proc. 4th IAU Symposium 119, Bangalore, 1985, Reidel,
  Dordrecht)\/}

\bibitem{Part}
{Parthasarathy} S, {Wagh} S~M and {Dadhich} N 1986 {\em Astrophys. J.\/} {\bf
  307} 38

\bibitem{dadhich12b}
{Dadhich} N 2012 {\em ArXiv e-prints\/} (\textit{Preprint} \eprint{1210.1041})

\bibitem{Koide14}
{Koide} S and {Baba} T 2014 {\em Astrophys J.\/} {\bf 792} 88
  (\textit{Preprint} \eprint{1407.7088})

\bibitem{Blandford1977}
{Blandford} R~D and {Znajek} R~L 1977 {\em Mon. Not. Roy. Astron. Soc.\/} {\bf
  179} 433--456

\bibitem{Banados09}
{Ba{\~n}ados} M, {Silk} J and {West} S~M 2009 {\em Physical Review Letters\/}
  {\bf 103} 111102

\bibitem{Virbhadra1}
{Virbhadra} K~S and {Ellis} G~F~R 2000 {\em Phys. Rev. D\/} {\bf 62} 084003

\bibitem{Virbhadra2}
{Virbhadra} K~S 2009 {\em Phys. Rev. D\/} {\bf 79} 083004

\bibitem{Bozza2005}
{Bozza} V, {de Luca} F, {Scarpetta} G and {Sereno} M 2005 {\em Phys. Rev. D.\/}
  {\bf 72} 083003

\bibitem{Bambi10}
{Bambi} C and {Yoshida} N 2010 {\em Classical and Quantum Gravity\/} {\bf 27}
  205006 (\textit{Preprint} \eprint{1004.3149})

\bibitem{Grenzebach2014}
{Grenzebach} A, {Perlick} V and {L{\"a}mmerzahl} C 2014 {\em Phys. Rev. D\/}
  {\bf 89} 124004 (\textit{Preprint} \eprint{1403.5234})

\bibitem{Falcke2013b}
{Falcke} H and {Markoff} S~B 2013 {\em Class. Quantum Grav\/} {\bf 30} 244003

\bibitem{Abdujabbarov15}
{Abdujabbarov} A~A, {Rezzolla} L and {Ahmedov} B~J 2015 {\em ArXiv e-prints\/}
  (\textit{Preprint} \eprint{1503.09054})

\bibitem{Stuchlik13}
{Stuchl{\'{\i}}k} Z and {Schee} J 2013 {\em Classical and Quantum Gravity\/}
  {\bf 30} 075012

\bibitem{Stuchlik10}
{Stuchl{\'{\i}}k} Z and {Schee} J 2010 {\em Classical and Quantum Gravity\/}
  {\bf 27} 215017 (\textit{Preprint} \eprint{1101.3569})

\bibitem{dadhich13}
{Dadhich} N, {Ghosh} S~G and {Jhingan} S 2013 {\em Phys. Rev. D\/} {\bf 88}
  124040 (\textit{Preprint} \eprint{1308.4770})

\bibitem{dadhich2013}
{Dadhich} N and {Ghosh} S~G 2013 {\em ArXiv e-prints\/} (\textit{Preprint}
  \eprint{1307.6166})

\bibitem{dadhich15}
{Dadhich} N 2015 {\em ArXiv e-prints\/} (\textit{Preprint} \eprint{1506.08764})

\bibitem{Hansen13}
{Hansen} D and {Yunes} N 2013 {\em Phys. Rev. D\/} {\bf 88} 104020
  (\textit{Preprint} \eprint{1308.6631})

\bibitem{dadhich13b}
{Dadhich} N 2013 {\em General Relativity and Gravitation\/} {\bf 45} 2383--2388
  (\textit{Preprint} \eprint{1301.5314})

\bibitem{frolov08}
{Frolov} V~P and {Kubiz{\v n}{\'a}k} D 2008 {\em Classical and Quantum
  Gravity\/} {\bf 25} 154005 (\textit{Preprint} \eprint{0802.0322})

\bibitem{Frolov03}
{Frolov} V and {Stojkovi{\'c}} D 2003 {\em Phys. Rev. D\/} {\bf 68} 064011
  (\textit{Preprint} \eprint{gr-qc/0301016})

\bibitem{Papnoi14}
{Papnoi} U, {Atamurotov} F, {Ghosh} S~G and {Ahmedov} B 2014 {\em Phys. Rev.
  D\/} {\bf 90} 024073 (\textit{Preprint} \eprint{1407.0834})

\bibitem{hackmann08}
{Hackmann} E, {Kagramanova} V, {Kunz} J and {L{\"a}mmerzahl} C 2008 {\em Phys.
  Rev. D\/} {\bf 78} 124018 (\textit{Preprint} \eprint{0812.2428})

\bibitem{Hioki09}
{Hioki} K and {Maeda} K~I 2009 {\em Phys. Rev. D\/} {\bf 80} 024042

\bibitem{Zaharov}
{Zakharov} A~F, {Nucita} A~A, {De Paolis} F and {Ingrosso} G 2005 {\em New
  Astron. Rev.\/} {\bf 10} 479--489

\bibitem{wei}
{Wei} S~W and {Liu} Y~X 2013 {\em JCAP\/} {\bf 11} 063

\bibitem{Vazquez}
{V{\'a}zquez} S~E and {Esteban} E~P 2004 {\em Nuovo Cim. B\/} {\bf 119} 489

\bibitem{Amarilla12}
{Amarilla} L and {Eiroa} E~F 2012 {\em Phys. Rev. D\/} {\bf 85} 064019

\bibitem{Atamurotov13b}
{Atamurotov} F, {Abdujabbarov} A and {Ahmedov} B 2013 {\em Phys. Rev. D\/} {\bf
  88} 064004

\bibitem{Stuchlik12a}
{Stuchl{\'{\i}}k} Z and {Schee} J 2012 {\em Classical and Quantum Gravity\/}
  {\bf 29} 065002

\bibitem{Amarilla10}
{Amarilla} L, {Eiroa} E~F and {Giribet} G 2010 {\em Phys. Rev. D\/} {\bf 81}
  124045

\bibitem{Amarilla13}
{Amarilla} L and {Eiroa} E~F 2013 {\em Phys. Rev. D\/} {\bf 87} 044057

\bibitem{chandra98}
Chandrasekhar S 1998 {\em {The mathematical theory of black holes}\/} (New
  York: Oxford University Press.)

\end{thebibliography}

\end{document}